# Constraints on the value of the fine structure constant from gravitational thermodynamics

**P.C.W. Davies**

The Beyond Center, Arizona State University, Tempe, AZ, U.S.A.

E-mail: deepthought@asu.edu

**Abstract**

In this paper I show how the second law of thermodynamics, generalized to include event horizon area, places interesting constraints on the value of the fine structure constant $\alpha$. A simple analysis leads to the approximate constraint $\alpha < 1$, although I suggest ways in which this bound might be reduced by a more detailed calculation. The analysis also leads to the conclusion that the existence of classical Dirac and point-like magnetic monopoles are inconsistent with the second law, and that GUT monopoles are inconsistent with either the second law or the existence of minicharged particles.

PACS numbers: 04.62.+v, 04.70Dy, 14.80.Hv, 98.80.Jk

The fine structure constant $\alpha = e^2/\hbar c \approx 1/137$ is one of the fundamental parameters of the standard model of particle physics. There is a long history of attempts to derive the measured value of α from an underlying theory, or exhibit it in the form of a compact mathematical expression (Boyer 1968, Chew 1981, Das and Coffman 1967, Dirac 1931, Eddington 1946, Rosen 1976, Ross 1986, Wyler 1969). The most significant advance in this endeavour was made by Dirac, who showed that if magnetic monopoles exist, with magnetic charge $\mu$, then

$$e\mu = n/2, \quad n = 1, 2, 3... \tag{1}$$

in units $\hbar = c = 1$ (which I use henceforth), from which it follows that

$$\mu^2 > 137/4. \tag{2}$$

The radius and area of the event horizon of a non-rotating black hole of mass *M*, electric charge *e* and magnetic charge *µ* is

$$r = M + \sqrt{[M^2 - e^2 - \mu^2]} \quad (3)$$

$$A = 4\pi r^2 \quad (4)$$

respectively, in units $G = 1$. It follows by inspection of Eqs. (3) and (4) that the presence of electric and magnetic charges serve to reduce the horizon area. The Bekenstein-Hawking entropy of the black hole is defined as

$$S = \tfrac{1}{4} A \quad (5)$$

in units $k = 1$, so at first glance it might seem as if dropping an electrically and/or magnetically charged particle into a black hole will reduce its entropy, in violation of the generalized second law of thermodynamics (Bekenstein 1973). However, the infalling particle will deliver mass as well as charge to the black hole, and the additional mass will increase the horizon area, thus opposing the effect of the charge.

Consider a point particle with rest mass $m$ and electric charge $e$ dropped from rest at infinity into a black hole of initial mass $M_i$ and radius $r_i = 2M_i$. The final horizon radius will then be

$$r_f = M_i + m + \sqrt{[(M_i + m)^2 - e^2]} \quad (6)$$

(I am neglecting the fact that some mass energy will be radiated electromagnetically by the infalling particle due to the nonlocal nature of the electromagnetic field and spacetime curvature effects. The radiation will reduce the final mass $M$ of the black hole somewhat, and render inequality (7) more stringent.) The generalized second law of thermodynamics then requires $r_f \geq r_i$, or

$$e^2 \leq 4M_i m. \quad (7)$$

Taken on its own, inequality (7) implies that one could achieve a violation of the second law merely by dropping a charged particle into a black hole of mass $M_i < e^2/4m$. However, this is to neglect the effects of quantum mechanics. If the diameter of the black hole is smaller than about the Compton wavelength $\lambda_c$ of the sacrificial particle, the particle will tend to scatter rather than enter the black hole. So to deliver the charge to the hole we require

$$1/m \lesssim 4M_i. \quad (8)$$

If we consider a black hole of the smallest mass consistent with (8), i.e.

$$M_i \approx 1/4m, \quad (9)$$

and demand that the generalized second law of thermodynamics *must* be satisfied, then combining (7) with (9) serves to provide a lowest upper bound on $e^2$:

$$e^2 \lesssim 1 \tag{10}$$

and hence (in the units being used)

$$\alpha \lesssim 1. \tag{11}$$

It may seem surprising that gravitational theory and thermodynamics can yield a bound on the fine structure constant of electrodynamics. On closer inspection, it is not so unexpected. It is well known (Zel'dovich and Popov 1970) that violating the bound renders the electromagnetic vacuum unstable, leading to pair creation. The thermodynamic properties of black holes stem from the Hawking effect, which is closely analogous: the gravitational field of the hole destabilizes the quantum vacuum in its vicinity, leading to particle creation and the outflow of Hawking radiation (Hawking 1975). In the case that the black hole also carries an electric charge, the vacuum destabilizing effect is augmented (Gibbons 1975). So there is a clear link. In a more realistic scenario one might also include the effect of the electron spin being delivered to the black hole, but this is suppressed by a factor $M/m_{\text{Planck}}$ compared to the charge, so I shall neglect it.

The inequality (10) is undeniably a crude estimate. A more detailed analysis of the propagation of a charged particle in the vicinity of a microscopic black hole might even turn out to yield a *lower* upper bound on the fine structure constant. Why might this be so? It is not sufficient, for example, to demand that the probability of the infalling particle to enter the black hole should merely be low; even an occasional conflict with the second law is unacceptable. Rather, the probability per unit time that the particle fails to scatter from the hole should be no greater than the probability per unit time of a spontaneous fluctuation of the black hole's horizon area from its equilibrium value of the same amount. The more detailed calculation required is beyond the scope of this paper.

In the case that the sacrificed particle is a magnetic monopole, the analysis is the same, but now the analogues of (8) and (11) are

$$\mu^2 \lesssim 4M_i m \tag{12}$$

$$\mu^2 \lesssim 1. \tag{13}$$

Clearly inequality (13) comes into conflict with inequality (2). The conflict is resolved if one requires that the monopole is not a point particle but an extended object, with a size a couple of orders of magnitude greater than its Compton wavelength, so that the minimum mass of a black hole that will swallow it needs to be correspondingly greater. Consider, for example, the GUT monopole of mass $10^{17}$ GeV and size ~ 1 fm; this object would clearly satisfy (2) and (12) with many orders of magnitude to spare. However, the classical Dirac monopole is a different case. For example, assume it has a size comparable to the classical electron radius $2.8 \times 10^{-15}$m $\backsim 10^{-3}\lambda_c$ (Giacomelli et. al. 2007), then its finite size will not be the major determining factor in its interaction with the black hole, and inequality (12) will still hold. Unless a more accurate calculation results in a bound one or two orders of magnitude less stringent than the above, then we may conclude that gravitational thermodynamic arguments combined with the Dirac quantization condition can be used to rule out both point-like and classical Dirac magnetic monopoles.

Note that, because of condition (1), an upper bound on the monopole charge implies a *lower* bound on the electric charge, and hence the fine structure constant, and vice versa. The symmetry between electric and magnetic charges in my analysis is broken only by invoking the finite size of the monopole. Let the monopole radius and mass be $\rho$ and $m_{\text{mono}}$ respectively. Then the requirement that $\rho > M$ combined with (12) gives a new fundamental bound

$$e^2 \gtrsim 1/\rho m_{\text{mono}}. \tag{14}$$

Recently attention has been given to the possible existence of minicharged particles (MCPs) as a possible explanation of dark matter. MCPs have been proposed with charges as low as $10^{-15}e$ (see, for example, Jaekel 2007). If this charge is used with inequality (14) one obtains

$$\rho \gtrsim 10^{30}/m_{\text{mono}} \tag{15}$$

For a monopole of mass $\backsim 10^{17}$GeV, this implies a minimum monopole size $\backsim 1$mm! Thus, unless one is prepared to entertain violations of the second law, MCPs are incompatible with the existence of microscopic magnetic monopoles, such as arise naturally in GUTs.

Finally, I shall consider the case of de Sitter horizons, which by general consent may be considered to have an associated temperature and entropy (Gibbons and Hawking 1977). Let a particle of mass $m$ and charge $e$ lie at the center of the static de Sitter coordinate system. The cosmological horizon radius $r_c$ is then the largest root of the quartic equation

$$1 - 2m/r + e^2/r^2 - \Lambda r^2 = 0 \tag{16}$$

where $\Lambda$ is $\frac{1}{3}$ of the cosmological constant (see, for example, Davies, Ford and Page 1986). For small $e$ and $m$, the largest root is $\backsim \Lambda^{-1/2}$, so in realistic scenarios $e^2\Lambda \ll 1$ and $m\Lambda^{1/2} \ll 1$. Then

$$r_c^2 \approx 1/\Lambda - 2m/\Lambda^{1/2} + e^2, \tag{17}$$

from which we see that the presence of the electric charge serves to *increase* the horizon area, unlike in the black hole case. An identical result follows for a magnetic monopole by replacing $e^2$ by $\mu^2$ (or for a dyon by replacing $e^2$ with $e^2 + \mu^2$). If the particle is displaced slightly from the center of coordinates (the observer remaining there), then the particle and observer will begin to separate under the action of the cosmological expansion. Eventually the particle will disappear over the observer's horizon. At late times, the observer's horizon area will be proportional to

$$r_c^2(t \to \infty) = 1/\Lambda \tag{18}$$

so the second law of thermodynamics generalized to cosmological horizons demands

$$e^2 \leq 2m\Lambda^{1/2} \tag{19}$$

Once again there is a quantum restriction on $m$: the particle must "fit" into the de Sitter space. A crude estimate is to demand that its Compton wavelength is less than the diameter of de Sitter space, which implies

$$m \gtrsim 2/\Lambda^{1/2} \tag{20}$$

leading to approximately the same bound (10) and (11) as in the black hole case. However, the de Sitter scenario is fundamentally different. The link between charge and vacuum instability is now tenuous in the extreme (de Sitter space doesn't radiate, i.e . the stress-energy-momentum tensor of the de Sitter vacuum does not correspond to that of thermal radiation). Consequently, there is no simple argument on quantum vacuum instability grounds, as in the black hole case. Moreover, it is possible to improve on the crude criterion (20) by considering the propagation of particles in de Sitter space. The analysis of quasi-normal modes of scalar and spinor fields in de Sitter space has been considered by Du, Wang and Su (2004), who find a bound on the *rest* mass for the scalar case:

$$m_0 > (d-1)/r_c \tag{21}$$

where $d$ is the spacetime dimensionality. For four dimensions, inequality (21) is the same order of magnitude as (20). However, the rest mass is not really the relevant quantity; rather, we need the total mass energy of the lowest energy quasi-normal mode that "fits" into de Ditter space. Du, Wand and Su find

$$[m_0^2 - \Lambda^{-1}]^{1/2} \tag{22}$$

for the real part of the lowest eigenmode (note there are no quasi-normal modes in the massless limit). If we interpret the mass-energy of the mode to be $\hbar$ times (22), then (19) should be replaced by the stronger inequality

$$e^2 \leq 2[m_0^2\Lambda - 1]^{1/2}. \tag{23}$$

For $m_0 \approx \Lambda^{-1/2}$, the upper bound on $e^2$ and hence $\alpha$ could be significantly $< 1$. Again, what is required is a more detailed calculation. This calculation needs to take into account the fact that $r_c$ in the above analysis has been calculated using the classical solution (16) of the Einstein-Maxwell equations in which the charged particle is localized at the origin. A more sophisticated calculation would be to use the expectation value of the stress energy momentum tensor of the lowest eigenmode in de Sitter space on the right hand side of Einstein's equations to compute the horizon radius.